\documentclass[showpacs,amsmath,amstex,amssymb,mathfonts,prl,twocolumn]{revtex4-1}
\usepackage{graphicx,bm,units}
\usepackage{amsmath,amssymb}
\usepackage{color, palatino,  wrapfig}

\newcommand{\be}{\begin{equation}}
\newcommand{\ee}{\end{equation}}

\newcommand{\ba}{\begin{eqnarray}}
\newcommand{\ea}{\end{eqnarray}}

\newcommand{\ignore}[1]{}
\newcommand{\pb}{{\textbf{p}}}

\begin{document}

\title{Absence of topological insulator phases in non-Hermitian PT-symmetric Hamiltonians}

\author{Yi Chen Hu}
\author{Taylor L. Hughes}
\affiliation{Department of Physics, University of Illinois, 1110 West Green St, Urbana IL 61801}
\begin{abstract}
In this work we consider a generalization of the symmetry classification of topological insulators to non-Hermitian Hamiltonians which satisfy a combined $PT$-symmetry (parity and time-reversal). We show via examples, and explicit bulk and boundary state proofs that the typical paradigm of forming topological insulator states from Dirac Hamiltonians is not compatible with the construction of non-Hermitian $PT$-symmetric Hamiltonians. The topological insulator states are $PT$-breaking phases and have energy spectra which are complex (not real) and thus are not consistent quantum theories. 
\end{abstract}
\date{\today}

\maketitle
With recent interest generated in the field of topological insulators and superconductors\cite{kane2005A,bernevig2006a,bernevig2006c,koenig2007, moore2007,fu2007b,roy2009a,hsieh2008}, the symmetry classification of (primarily free-fermion) Hamiltonians has resurfaced\cite{dyson1962,altland1997,schnyder2008A,qi2008B,kitaev2009}. The first example of a topological insulator (TI), the integer quantum Hall effect (IQHE), is gapped in the bulk and exhibits gapless, robust states on its boundaries. Most other examples of topological insulators\cite{kane2005A,bernevig2006c,koenig2007, moore2007,fu2007b,roy2009a,hsieh2008} share these characteristic features. One interesting distinction between the IQHE and, for example, the quantum spin Hall effect (QSHE), is that the IQHE is  completely robust to any type of Hamiltonian perturbation, while the QSHE is only robust to perturbations that preserve time-reversal symmetry $T.$ A full symmetry protected classification of topological insulators\cite{qi2008B,schnyder2008A} and superconductors\cite{schnyder2008A} based on charge-conjugation, time-reversal, and chiral symmetries was then unified into a periodic table\cite{kitaev2009}. 

In addition to the classification theory of  TI's, Hamiltonian symmetries are important in the theory of non-Hermitian Hamiltonians\cite{bender1998,bender2007,bender2010,jonessmith2010} where it has been shown that non-Hermitian Hamiltonians can still describe viable quantum systems as long as $PT$ symmetry is unbroken. By $PT$ we mean the combined operation of a parity/inversion symmetry $P,$ and $T.$ Given a non-Hermitian Hamiltonian $H$ that satisfies $[H,PT]=0$ one can provide necessary and sufficient conditions that the energy spectrum of $H$ be real\cite{bender2010}. It is thus natural to attempt to extend the periodic table of TI's to non-Hermitian Hamiltonians which satisfy $PT$-symmetry. Although our initial hope was to find non-Hermitian examples of TI states we instead show that TI phases are incompatible with the $PT$-symmetric construction of (at least a large class of) non-Hermitian Hamiltonians. The TI states are $PT$-breaking and exhibit imaginary eigenvalues even when $H$ has been constructed to preserve $PT$-symmetry. 

In this Letter we start by showing a few pedagogical examples of PT-symmetric Dirac Hamiltonians that are perturbed away from being Hermitian. We offer these examples to show how the construction of non-Hermitian $PT$-symmetric  TI states fails. Our focus is on Dirac Hamiltonians because they are the minimal models for topological insulators. An extension to generic insulator Hamiltonians is straight-forward. After the examples, we provide a more generic proof of necessary conditions for the bulk spectra of such Dirac Hamiltonians to have a fully real spectrum (\emph{i.e.} eigenvalues are real for all values of the momenta). Finally, we give some arguments about the properties of the gapless boundary states that show that PT-symmetry, non-Hermiticity, and topological insulator states do not seem to be compatible. 

Before we begin with the examples let us list the relevant symmetry properties we will use in this Letter. The $T$-operator is represented by $T=UK$ where $U$ is a unitary operator and $K$ is complex conjugation. Depending on $U$ we can have $T^2=\pm 1.$ For a Bloch Hamiltonian $H(p)$ to be invariant under $T$ we must have $TH(\pb)T^{-1}=H(-\pb).$ The $P$-symmetry, which we will call \emph{parity} when needed, is a unitary operator with $P^2=+1.$ There is no requirement on which spatial coordinates the $P$ operator inverts and for now we will leave it generic. The condition that a Bloch Hamiltonian be $P$-invariant is $PH(\pb)P^{-1}=H(\bar{\pb})$ where $\bar{\pb}$ is a symbol  characterizing a given $P$ by indicating  which coordinates are inverted and which remain unaffected. As an example, if all the coordinates are inverted then $\bar{\pb}=-\pb.$ Finally, the condition that a Bloch Hamiltonian be $PT$-symmetric is $PTH(\pb)(PT)^{-1}=H(-\bar{\pb})$ and equivalently requires that $H(\pb)$ is either both odd or both even under $P$ and $T$ separately. We occasionally mention charge-conjugation symmetry $C$ which requires $CH(\pb)C^{-1}=-H^{\ast}(-\pb).$

\section{1D Dirac Hamiltonians}
  \begin{figure}[t!]
\includegraphics[width=0.45\textwidth]{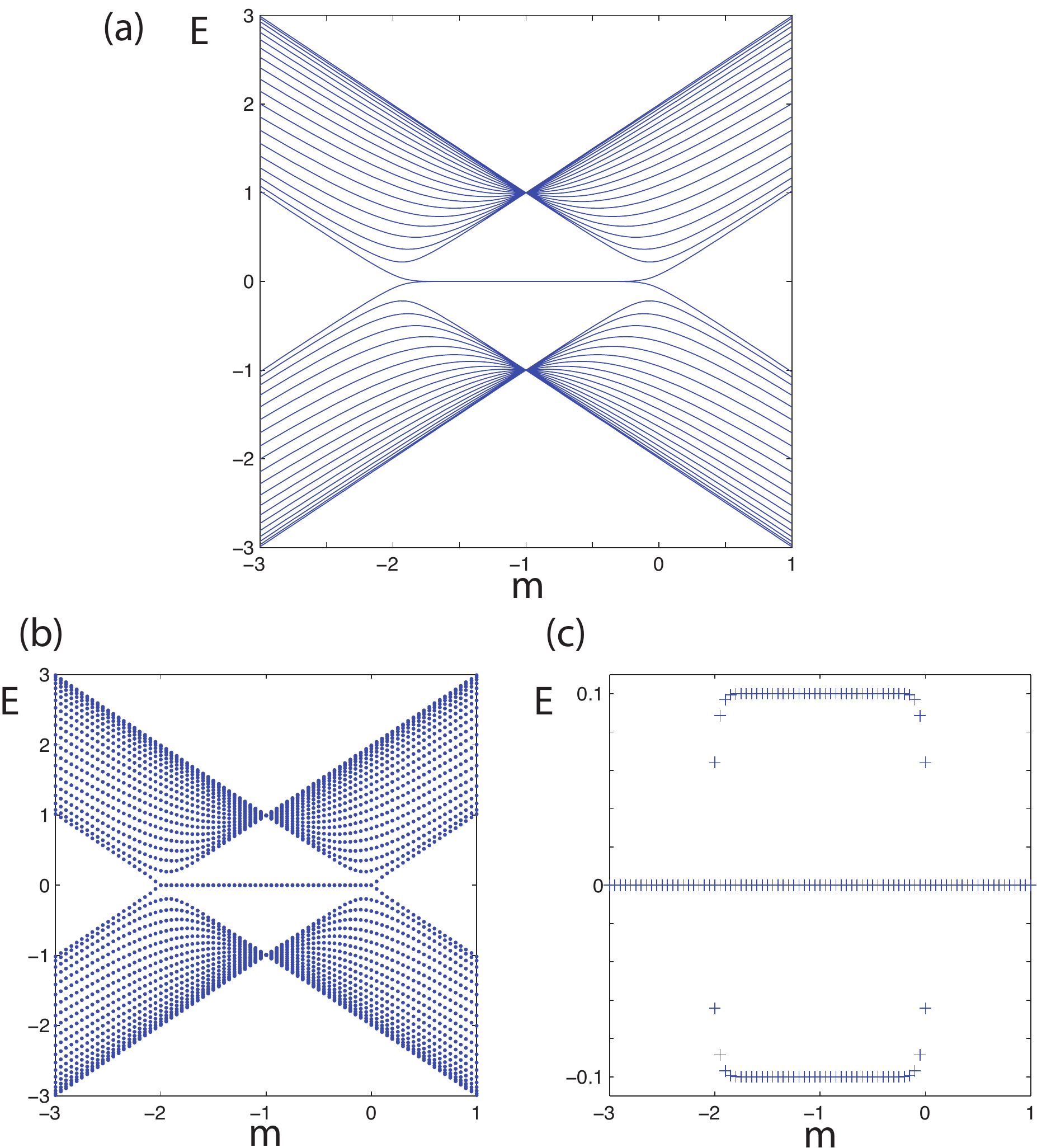}\\
\caption{Energy spectrum of $H'_{1D,lattice}$ (Eq. \ref{eq:H1Dlattice}) with open boundaries vs. $m$ (a)$\Delta=0$ gives a fully real spectrum with zero-modes for $-2<m<0.$(b) real part of energy spectrum and (c) imaginary part of energy spectrum for  $\Delta=0.1.$ Complex eigenvalues exist in the former topological phase.}\label{fig:1}
\end{figure}

We will begin the examples in 1D with the gapped continuum Dirac Hamiltonian
\begin{equation}
H_{1D}(p)= v_F p\sigma^y+m\sigma^z\label{eq:1DHA}
\end{equation}\noindent which is P, T, and PT symmetric with the symmetry operators $T=K,$ $P=\sigma^z$ and $\bar{p}=-p.$ This Hamiltonian also satisfies a $C$ symmetry with $C=\sigma^x$ and provides an example of a $Z_2$ topological insulator protected by $C$ symmetry\cite{qi2008B}. 
 We can add an additional term to the Hamiltonian to get
\begin{equation}
H^{'}_{1D}(p)= p\sigma^y+m\sigma^z+i\Delta\sigma^x\label{eq:1DHB}
\end{equation}\noindent which is P-odd, T-odd, C-even, PT-even and non-Hermitian and where we set $v_F=1.$  Since we have not broken $C$-symmetry the $Z_2$ classification naively should remain intact. 1D Dirac Hamiltonians with non-Hermitian potentials have been studied in, for example, Ref. \onlinecite{sinha2005}. The energy spectrum of this Hamiltonian is simple to calculate: 
\begin{eqnarray}
E_{\pm}=\pm\sqrt{p^2+m^2-\Delta^2}
\end{eqnarray}\noindent which is real as long as $\vert m\vert>\vert\Delta\vert.$  So we see it is possible to add a non-Hermitian perturbation to the Dirac Hamiltonian and keep the spectrum to be entirely real for values of $m$ both $<0$ and $>0.$ Note that the perturbation has the special property $\{i\Delta\sigma^x,H_{1D}(p)\}=0$ which will become important later. 

Although it is promising that there is a regime where this non-Hermitian Hamiltonian will have a real spectrum, there is already something worrying about the spectrum. With $\Delta=0$ this model has a gap-closing phase transition at $m=0$ which separates a trivial insulator phase from a topological insulator phase\cite{qi2008B}. We see here that if $\Delta\neq 0$ this phase transition becomes destabilized when $0<\vert m\vert <\vert \Delta\vert.$ To properly describe the TI phase we need to use a lattice version of this Dirac Hamiltonian with a Bloch form
\begin{equation}
H'_{1D,lattice}(p)=(\sin p) \sigma^y+(1+m-\cos p)\sigma^z +i\Delta\sigma^x\label{eq:H1Dlattice}
\end{equation}\noindent where we have set the lattice constant $a=1.$ When $\Delta=0$ the lattice Hamiltonian is in a trivial insulator phase when $m<-2$ or $m>0,$ and a topological insulator when $-2<m<0.$ With non-vanishing $\Delta$ we have $E_{\pm}=\pm\sqrt{1+(1+m)^2-\Delta^2-2(1+m)\cos p}.$ To be a viable spectrum this must be real for all $-\pi\leq p \leq \pi.$ In Fig. \ref{fig:1}a we show the spectrum for $\Delta=0$ as a function of $m$ with open boundary conditions. This clearly shows the boundary state zero modes which persist when $-2<m<0.$ In Fig. \ref{fig:1}b,c we show the real and imaginary parts of the energy spectrum when $\Delta=0.1.$ This shows that as soon as the system nears the phase boundary to the topological insulator state the spectrum develops imaginary pieces and thus generates a $PT$-breaking phase. This is a common feature and shows that the 1D topological insulator state here cannot be reached when $\Delta\neq 0.$

\section{2D Dirac Hamiltonians}
2D becomes more complicated because of two reasons (i) there are two natural definitions of $P$: parity $(x,y)\to (x,-y)$ and inversion $(x,y)\to(-x,-y)$ which both satisfy $P^2=1$ (ii) from the conventional classification theory of topological insulators it is natural to look at both $2$-band and $4$-band models. We won't exhaust all these cases, but only provide some instructive examples.  

\emph{2-band models:} Let us consider the Hamiltonian which represents the continuum model of a Chern insulator\cite{haldane1988}:
\begin{equation}
H^{(2)}_{2D}(\pb)=p_x\sigma^x+p_y\sigma^y+m\sigma^z
\end{equation}\noindent with $T=K$ and $P=\sigma^x.$ This $P$ corresponds to a parity symmetry and sends $p_y\to -p_y$ so $\bar{\pb}=(p_x,-p_y).$ Unfortunately this Hamiltonian cannot be made $PT$-symmetric  even before we perturb it because the mass term breaks $PT.$
Next, let us consider the same Hamiltonian with $T=K$ and $P=\sigma^z$ where this $P$ is an inversion symmetry with $\bar{\pb}=-\pb.$ Again even the base Hamiltonian is not $PT$-symmetric because the $p_x$ term breaks $PT.$
Finally, if we consider the same Hamiltonian with $T=i\sigma^y K$ and $P=\sigma^z$ and $\bar{\pb}=-\pb$ the mass term breaks $PT.$ 
From these few attempts it seems like we cannot get any interesting non-Hermitian TI Hamiltonians with 2-bands. We will see why this is so in the general proof section, but the basic idea is that there are no additional matrices $M$ which anticommute with $H^{(2)}_{2D}(\pb).$ 

\emph{4-band models:} Let us start with a continuum QSH Hamiltonian\cite{kane2005A,bernevig2006c}:
\begin{equation}
H^{(4)}_{2D}(\pb)=p_x\Gamma^1+p_y\Gamma^2+m\Gamma^0
\end{equation}\noindent with $\Gamma^1=\tau^x\otimes\sigma^x,\; \Gamma^2=\tau^x\otimes\sigma^y,\;$  $\Gamma^3=\tau^y\otimes \mathbb{I},\;$ $\Gamma^4=\tau^x\otimes \sigma^z,\;$  and $\Gamma^0=\tau^z\otimes\mathbb{I}.$ This  has symmetry generators $T=i\sigma^y K$ and $P=\Gamma^0$ and $\bar{\pb}=-\pb.$  We can perturb this Hamiltonian with the non-Hermitian PT-symmetric matrices $\{i\tau^x\otimes \mathbb{I},\; i\tau^y\otimes\sigma^i,\; i\tau^z\otimes \sigma^i,\; i\mathbb{I}\otimes \sigma^i\}$
but none of these anti-commute with the Hamiltonian and will lead to imaginary bulk eigenvalues as we will see in the following section.

We can write down another Hamiltonian in 2D
\begin{equation}
H^{(4)}_{2DB}(p)=p_x\Gamma^2+p_y\Gamma^3+m\Gamma^0\label{eq:2DHA}
\end{equation}
This has symmetry operators $T=K$ and $P=\Gamma^0$ with $\bar{\pb}=-\pb.$ There are two interesting terms with which we can perturb the Hamiltonian
\begin{equation}
\Delta H^{(4)}_{2D}=i\Delta_1\Gamma^1+i\Delta_4\Gamma^4
\end{equation}\noindent  Adding these terms to the Hamiltonian gives us an energy spectrum
\begin{equation}
E_{\pm}=\pm \sqrt{p_{x}^2+p_{y}^2+m^2-\Delta_{1}^2-\Delta_{4}^{2}}
\end{equation}\noindent which is real as long as $m^2\geq \Delta_{3}^2+\Delta_{4}^2.$  However, this model, with $T=K$ is not known to exhibit a robust topological insulator state anyway, but we see there is the same problem with the gap-closing transition leading to a $PT$-breaking region.

\section{Necessary conditions for real bulk eigenvalues}
The generic features of the example models is that if there is a $PT$-symmetric, non-Hermitian matrix that anticommutes with $H(p)$ then we can have real eigenvalues, but the topological insulator phase is $PT$-breaking. If the only $PT$-symmetric non-Hermitian matrices commute with at least one term of $H(p)$ then even the bulk eigenvalues will not be real for all of momentum space. We formalize this statement now.  

{\textbf{Theorem:}} For a $PT$ symmetric Hamiltonian of the Dirac form $H=p\Gamma+mQ+i\Delta S$ where $\{Q,\Gamma\}=0$ and $S,Q,\Gamma$ are matrices with $S^2=Q^2=\Gamma^2=1$, the quantities $\{Q,S\}$ \emph{and} $\{\Gamma,S\}$ must vanish to have a real bulk spectrum. 

\emph{Proof:} Assume we have a Hamiltonian 
\begin{equation}
H(p)=p\Gamma+mQ+i\Delta S.
\end{equation}\noindent By contradiction let us assume that $[S,Q]=0$ so that the anti-Hermitian term commutes with the \emph{mass} term. For the point $p=0$ the Hamiltonian reduces to $H(0)=mQ+i\Delta S.$ Since $S$ and  $Q$ commute they can be simultaneously diagonalized. Since $S^2=Q^2=+1$ the eigenvalues of these matrices are $\pm 1.$ Thus the eigenvalues of $H(0)$ can only be $\pm m \pm i\Delta$ and are always imaginary if $\Delta\neq 0.$ This is a contradiction and so we know that $\{S,Q\}=0$ must hold if the spectrum is to be real for all allowed $p.$

Now we want to prove the second necessary condition by contradiction. Assume that $\{S,Q\}=0,$ which we now know must hold, but $[S,\Gamma]=0.$ The Hamiltonian at finite $p$ is as above and $H^2(p)=(p^2+m^2-\Delta^2)\mathbb{I}+ip\Delta\{\Gamma,S\}.$ From our assumptions $\{\Gamma,S\}=2\Gamma S$ and $(\Gamma S)^2=+1$ so the eigenvalues of $\{\Gamma,S\}$ are $\pm 1.$ Thus the eigenvalues of $H^2(p)$ are
\begin{eqnarray}
E^{2}_{\pm}&=&(p^2+m^2-\Delta^2)\pm 2ip\Delta\nonumber\\
&\implies & \pm E_{\pm}=\pm\sqrt{(p\pm i \Delta)^2+m^2}.
\end{eqnarray}\noindent This means that for all non-zero $p$ the eigenvalues will be imaginary. This is a contradiction and thus we have proven that both $\{S,Q\}=\{S,\Gamma\}=0$ are necessary conditions for the bulk spectra to be real. $\blacksquare$ 

{\textbf{Corollary:}} For generic Hamiltonians of the form $H(p)=p_a\Gamma^a+mQ +i\Delta S$ where $\{\Gamma^a,Q\}=0$ and $(\Gamma^{a})^2=Q^2=S^2=1$ we must have $\{S,\Gamma^a\}=\{S,Q\}=0$ for all values of $a.$

\emph{Proof:} The case involving the mass term is unchanged from above. For the momentum term, by contradiction, assume that there is a value $a=a_0$ such that $[S,\Gamma^{a_0}]=0.$ We can then set all $p_a=0$ for $a\neq a_0$ and use the same theorem as above to prove the result. $\blacksquare$

Thus, we see that to give a real spectrum we must have an $S$ that is a Clifford algebra generator along with the generators $\Gamma^a$ and $Q.$ To satisfy $PT$ symmetry we need
\begin{eqnarray}
(PT)iS(PT)^{-1}&=&-i(PU)S^{\ast}U^{-1}P^{-1}=iS\nonumber\\
&\implies & (PU)S^{\ast}+S(PU)=0
\end{eqnarray}\noindent this means that if $S=S^{\ast}$ then $PT$ symmetry requires $\{PU,S\}=0.$ If $S=-S^{\ast}$ then $PT$ symmetry requires $[PU,S]=0.$ The generic energy spectrum for such an anti-commuting $S$ will be $E_{\pm}=\pm\sqrt{\sum_{a}p_{a}^2+m^2-\Delta^2}.$

\section{Boundary/Interface States}
 Let us return back to the 1D Hamiltonian specified in Eqs. \ref{eq:1DHA},\ref{eq:1DHB}. There is a region of the TI phase (\emph{i.e.} where $-2<m<0$) where $\vert m\vert$ could be greater than $\vert\Delta\vert$ and thus could have a real bulk spectrum, but from Fig. \ref{fig:1}b,c we see that the phase is still $PT$-breaking. To illustrate why the TI state is $PT$-breaking we need to consider the gapless boundary states. We will do this in the simplest possible way and capture the essential details by considering two interfaces between regions described by Dirac Hamiltonians with $m(x)=-m_0$ for $x<-x_0$ next to a region with $m(x)=m_0>0$ for $-x_0<x<x_0$ and finally with $m(x)=-m_0$ for $x>x_0.$ We assume that $\vert m_0\vert\geq \vert\Delta\vert$ so that the bulk energy spectrum is real in each region and that $2x_0\gg \hbar v_F/m_0$ so that for our purposes the interfaces are independent of each other.  Note that this potential  satisfies $m(x)=m^{\ast}(-x)$ and is thus $PT$ symmetric. The $1D$ Hamiltonian is
\begin{equation}
H=-i\frac{d}{dx}\sigma^y+m(x)\sigma^z.
\end{equation}\noindent  This Hamiltonian has two zero-energy boundstates\cite{Jackiw1976} one per interface domain wall given by 
\begin{equation}
\Psi^{\pm}_0(x)=\frac{1}{\sqrt{2}}\exp\left[\pm\int_{\pm x_0}^{x}m(x')dx'\right]\left(\begin{array}{c}1\\ \mp 1\end{array}\right).
\end{equation}
We see something interesting that occurs here,
\begin{eqnarray}
PT\Psi^{+}_0(x)=P\Psi^{+ \ast}_0(x)=\sigma^z\Psi^{+\ast}_{0}(-x)=\Psi^{-}_{0}(x)
\end{eqnarray}\noindent This means that $\Psi_0$ is not an eigenstate of $PT$ and thus we do not expect it to have a real eigenvalue if we add a non-Hermitian $PT$ symmetric term\cite{bender2007,bender2010}. The $PT$ symmetry transforms one boundary state into the other one. This is not so strange since the $P$ symmetry should interchange the two ends of the $1D$ system, or in this case the states on each domain wall.  

Now let us perturb the Hamiltonian by $i\Delta\sigma^x,$ as in the 1D example above, and focus near a single domain wall. The unperturbed system has $\Psi_0$ as a zero energy state. Adding the perturbation is trivial since $\Psi_0$ is \emph{already} an eigenstate of $\sigma^x.$ Thus we see that perturbing the domain wall states changes the energies of the states from $E=0$ to $E=\pm i \Delta$ which is obviously imaginary. Thus, the domain wall, and by analogy, TI boundary states break the $PT$-symmetry and are not compatible with a real energy spectrum. 

To show that this is not a pathological case for the 1D Hamiltonian let us consider the  2D Hamiltonian in Eq. \ref{eq:2DHA} which will still have (unprotected) bound states on mass domain walls.  We will assume the domain walls are in the y-direction this time. The Hamiltonian is then
\begin{equation}
H^{(4)}_{2D}=p_x\Gamma^2-i\frac{d}{dy}\Gamma^3+m(y)\Gamma^0
\end{equation}\noindent where $p_x$ is a number. For $p_x=0$ we can use the same ansatz as in the 1D case but this  time there are two zero-energy solutions per wall. For an upward stepping domain wall we find
\begin{eqnarray}
\Psi_{0A}=\frac{1}{\sqrt{2}}\left(\begin{array}{c}1\\0\\-1\\0\end{array}\right),\;\;\;\;
\Psi_{0B}=\frac{1}{\sqrt{2}}\left(\begin{array}{c}0\\1\\0\\-1\end{array}\right).
\end{eqnarray}\noindent For $p_x\neq 0$ these solutions \emph{do not} automatically diagonalize $\Gamma^1$ and we are left with a reduced $2\times 2$ problem
\begin{eqnarray}
H^{(eff)}_{ij}&=&\langle\Psi_{0i}\vert p_x\Gamma^2\vert\Psi_{0j}\rangle=p_x\sigma^{y}_{ij}
\end{eqnarray}\noindent where $i,j=A,B.$ Now if we add the allowed, anti-commuting non-Hermitian terms $i\Delta_1\Gamma^1+i\Delta_4\Gamma^4$ the effective Hamiltonian of the edge states becomes
\begin{eqnarray}
H^{(eff)}_{ij}&=&p_x\sigma^{y}_{ij}+\langle\Psi_{0i}\vert i\Delta_1\Gamma^1+i\Delta_4\Gamma^4\vert\Psi_{0j}\rangle\nonumber\\
&=&p_x\sigma^{y}_{ij}+i\Delta_1\sigma^{x}_{ij}+i\Delta_4\sigma^{z}_{ij}
\end{eqnarray}\noindent This Hamiltonian is simple to diagonalize has energies
\begin{eqnarray}
E_{\pm}=\pm\sqrt{p_{x}^2-\Delta_{1}^2-\Delta_{4}^2}
\end{eqnarray}\noindent which are imaginary at least at $p_x=0$ if either of the $\Delta_i\neq 0.$ Thus again the boundary spectrum is imaginary. The states $\Psi_{0A/B}$ are not eigenstates of the $PT$ operator and will be transformed into the allowed states on the opposite wall by $PT.$ These arguments generalize to the other topological insulator classes and show that such a phase is generically $PT$-breaking. 

With this proof in place one can immediately begin to search for exceptions. The first place to look would be topological superconductors whose boundary states have weight on both boundaries. We performed a cursory test of some classes of topological superconductors and were not able to construct an interesting non-Hermitian PT-symmetric phase. The presence of Majorana boundary fermions which are non-local may provide a way around the boundary state problem but we leave this open for future work. Also, our result does not immediately apply to topological insulator states protected by point-group symmetries like inversion\cite{turner2010A,hughes2011,turner2010B} or even $C4$ symmetry\cite{fu2011}. In this case, however, we believe that our result can easily be adapted to at least a large class of these models since again they can usually expressed using minimal Dirac-type models where our results would immediately carry over. This does leave open the possibility of finding non-Hermitian topological phases, but without Dirac-type Hamiltonians at their foundation.

\emph{Acknowledgements} TLH thanks C. M. Bender for useful conversations. TLH is supported by NSF under grant DMR 0758462 at the University of Illinois, and by the ICMT.
\bibliography{TI-1}

\begin{thebibliography}{10}%
\makeatletter
\providecommand \@ifxundefined [1]{%
 \ifx #1\undefined \expandafter \@firstoftwo
 \else \expandafter \@secondoftwo
\fi
}%
\providecommand \@ifnum [1]{%
 \ifnum #1\expandafter \@firstoftwo
 \else \expandafter \@secondoftwo
\fi
}%
\providecommand \enquote [1]{``#1''}%
\providecommand \bibnamefont  [1]{#1}%
\providecommand \bibfnamefont [1]{#1}%
\providecommand \citenamefont [1]{#1}%
\providecommand\href[0]{\@sanitize\@href}%
\providecommand\@href[1]{\endgroup\@@startlink{#1}\endgroup\@@href}%
\providecommand\@@href[1]{#1\@@endlink}%
\providecommand \@sanitize [0]{\begingroup\catcode`\&12\catcode`\#12\relax}%
\@ifxundefined \pdfoutput {\@firstoftwo}{%
 \@ifnum{\z@=\pdfoutput}{\@firstoftwo}{\@secondoftwo}%
}{%
 \providecommand\@@startlink[1]{\leavevmode\special{html:<a href="#1">}}%
 \providecommand\@@endlink[0]{\special{html:</a>}}%
}{%
 \providecommand\@@startlink[1]{%
  \leavevmode
  \pdfstartlink
   attr{/Border[0 0 1 ]/H/I/C[0 1 1]}%
   user{/Subtype/Link/A<</Type/Action/S/URI/URI(#1)>>}%
  \relax
 }%
 \providecommand\@@endlink[0]{\pdfendlink}%
}%
\providecommand \url  [0]{\begingroup\@sanitize \@url }%
\providecommand \@url [1]{\endgroup\@href {#1}{\urlprefix}}%
\providecommand \urlprefix [0]{URL }%
\providecommand \Eprint[0]{\href }%
\@ifxundefined \urlstyle {%
  \providecommand \doi [1]{doi:\discretionary{}{}{}#1}%
}{%
  \providecommand \doi [0]{doi:\discretionary{}{}{}\begingroup
  \urlstyle{rm}\Url }%
}%
\providecommand \doibase [0]{http://dx.doi.org/}%
\providecommand \Doi[1]{\href{\doibase#1}}%
\providecommand \bibAnnote [3]{%
  \BibitemShut{#1}%
  \begin{quotation}\noindent
    \textsc{Key:}\ #2\\\textsc{Annotation:}\ #3%
  \end{quotation}%
}%
\providecommand \bibAnnoteFile [2]{%
  \IfFileExists{#2}{\bibAnnote {#1} {#2} {\input{#2}}}{}%
}%
\providecommand \typeout [0]{\immediate \write \m@ne }%
\providecommand \selectlanguage [0]{\@gobble}%
\providecommand \bibinfo [0]{\@secondoftwo}%
\providecommand \bibfield [0]{\@secondoftwo}%
\providecommand \translation [1]{[#1]}%
\providecommand \BibitemOpen[0]{}%
\providecommand \bibitemStop [0]{}%
\providecommand \bibitemNoStop [0]{.\EOS\space}%
\providecommand \EOS [0]{\spacefactor3000\relax}%
\providecommand \BibitemShut [1]{\csname bibitem#1\endcsname}%
\bibitem{kane2005A}%
  \BibitemOpen
  \bibfield{author}{%
  \bibinfo {author} {\bibnamefont{\textrm{C. L. Kane}}}\ and\ \bibinfo {author}
  {\bibnamefont{\textrm{E. J. Mele}}},\ }%
  \bibfield{journal}{%
  \bibinfo {journal} {Phys. Rev. Lett.}\ }%
  \textbf{\bibinfo {volume} {95}},\ \bibinfo {pages} {226801} (\bibinfo {year}
  {2005})%
  \bibAnnoteFile{NoStop}{kane2005A}%
\bibitem{bernevig2006a}%
  \BibitemOpen
  \bibfield{author}{%
  \bibinfo {author} {\bibnamefont{\textrm{B.A. Bernevig}}}\ and\ \bibinfo
  {author} {\bibnamefont{\textrm{S.C. Zhang}}},\ }%
  \bibfield{journal}{%
  \bibinfo {journal} {Phys. Rev. Lett.}\ }%
  \textbf{\bibinfo {volume} {96}},\ \bibinfo {pages} {106802} (\bibinfo {year}
  {2006})%
  \bibAnnoteFile{NoStop}{bernevig2006a}%
\bibitem{bernevig2006c}%
  \BibitemOpen
  \bibfield{author}{%
  \bibinfo {author} {\bibnamefont{\textrm{B. A. Bernevig}}}, \bibinfo {author}
  {\bibnamefont{\textrm{T. L. Hughes}}},\ and\ \bibinfo {author}
  {\bibnamefont{\textrm{S.C. Zhang}}},\ }%
  \bibfield{journal}{%
  \bibinfo {journal} {Science}\ }%
  \textbf{\bibinfo {volume} {314}},\ \bibinfo {pages} {1757} (\bibinfo {year}
  {2006})%
  \bibAnnoteFile{NoStop}{bernevig2006c}%
\bibitem{koenig2007}%
  \BibitemOpen
  \bibfield{author}{%
  \bibinfo {author} {\bibfnamefont{M.}~\bibnamefont{K\"onig}}, \bibinfo
  {author} {\bibfnamefont{S.}~\bibnamefont{Wiedmann}}, \bibinfo {author}
  {\bibfnamefont{C.}~\bibnamefont{Br\"une}}, \bibinfo {author}
  {\bibfnamefont{A.}~\bibnamefont{Roth}}, \bibinfo {author}
  {\bibfnamefont{H.}~\bibnamefont{Buhmann}}, \bibinfo {author}
  {\bibfnamefont{L.}~\bibnamefont{Molenkamp}}, \bibinfo {author}
  {\bibfnamefont{X.-L.}\ \bibnamefont{Qi}},\ and\ \bibinfo {author}
  {\bibfnamefont{S.-C.}\ \bibnamefont{Zhang}},\ }%
  \bibfield{journal}{%
  \bibinfo {journal} {Science}\ }%
  \textbf{\bibinfo {volume} {318}},\ \bibinfo {pages} {766} (\bibinfo {year}
  {2007})%
  \bibAnnoteFile{NoStop}{koenig2007}%
\bibitem{moore2007}%
  \BibitemOpen
  \bibfield{author}{%
  \bibinfo {author} {\bibfnamefont{J.~E.}\ \bibnamefont{Moore}}\ and\ \bibinfo
  {author} {\bibfnamefont{L.}~\bibnamefont{Balents}},\ }%
  \bibfield{journal}{%
  \Doi{10.1103/PhysRevB.75.121306}{\bibinfo {journal} {Phys. Rev. B}}\ }%
  \textbf{\bibinfo {volume} {75}},\ \bibinfo {eid} {121306} (\bibinfo {year}
  {2007})%
  \bibAnnoteFile{NoStop}{moore2007}%
\bibitem{fu2007b}%
  \BibitemOpen
  \bibfield{author}{%
  \bibinfo {author} {\bibfnamefont{L.}~\bibnamefont{Fu}}, \bibinfo {author}
  {\bibfnamefont{C.~L.}\ \bibnamefont{Kane}},\ and\ \bibinfo {author}
  {\bibfnamefont{E.~J.}\ \bibnamefont{Mele}},\ }%
  \bibfield{journal}{%
  \Doi{10.1103/PhysRevLett.98.106803}{\bibinfo {journal} {Phys. Rev. Lett.}}\
  }%
  \textbf{\bibinfo {volume} {98}},\ \bibinfo {eid} {106803} (\bibinfo {year}
  {2007})%
  \bibAnnoteFile{NoStop}{fu2007b}%
\bibitem{roy2009a}%
  \BibitemOpen
  \bibfield{author}{%
  \bibinfo {author} {\bibfnamefont{R.}~\bibnamefont{Roy}},\ }%
  \bibfield{journal}{%
  \bibinfo {journal} {Phys. Rev. B}\ }%
  \textbf{\bibinfo {volume} {79}},\ \bibinfo {pages} {195322} (\bibinfo {year}
  {2009})%
  \bibAnnoteFile{NoStop}{roy2009a}%
\bibitem{hsieh2008}%
  \BibitemOpen
  \bibfield{author}{%
  \bibinfo {author} {\bibfnamefont{D.}~\bibnamefont{Hsieh}}, \bibinfo {author}
  {\bibfnamefont{D.}~\bibnamefont{Qian}}, \bibinfo {author}
  {\bibfnamefont{L.}~\bibnamefont{Wray}}, \bibinfo {author}
  {\bibfnamefont{Y.}~\bibnamefont{Xia}}, \bibinfo {author}
  {\bibfnamefont{Y.~S.}\ \bibnamefont{Hor}}, \bibinfo {author}
  {\bibfnamefont{R.~J.}\ \bibnamefont{Cava}},\ and\ \bibinfo {author}
  {\bibfnamefont{M.~Z.}\ \bibnamefont{Hasan}},\ }%
  \bibfield{journal}{%
  \bibinfo {journal} {Nature}\ }%
  \textbf{\bibinfo {volume} {452}},\ \bibinfo {pages} {970} (\bibinfo {year}
  {2008})%
  \bibAnnoteFile{NoStop}{hsieh2008}%
\bibitem{dyson1962}%
  \BibitemOpen
  \bibfield{author}{%
  \bibinfo {author} {\bibfnamefont{F.~J.}\ \bibnamefont{Dyson}},\ }%
  \bibfield{journal}{%
  \bibinfo {journal} {J. Math. Phys.}\ }%
  \textbf{\bibinfo {volume} {3}},\ \bibinfo {pages} {1199} (\bibinfo {year}
  {1962})%
  \bibAnnoteFile{NoStop}{dyson1962}%
\bibitem{altland1997}%
  \BibitemOpen
  \bibfield{author}{%
  \bibinfo {author} {\bibfnamefont{A.}~\bibnamefont{Altland}}\ and\ \bibinfo
  {author} {\bibfnamefont{M.~R.}\ \bibnamefont{Zirnbauer}},\ }%
  \bibfield{journal}{%
  \bibinfo {journal} {Phys. Rev. B}\ }%
  \textbf{\bibinfo {volume} {55}},\ \bibinfo {pages} {1142} (\bibinfo {year}
  {1997})%
  \bibAnnoteFile{NoStop}{altland1997}%
\bibitem{schnyder2008A}%
  \BibitemOpen
  \bibfield{author}{%
  \bibinfo {author} {\bibfnamefont{A.~P.}\ \bibnamefont{Schnyder}}, \bibinfo
  {author} {\bibfnamefont{S.}~\bibnamefont{Ryu}}, \bibinfo {author}
  {\bibfnamefont{A.}~\bibnamefont{Furusaki}},\ and\ \bibinfo {author}
  {\bibfnamefont{A.~W.~W.}\ \bibnamefont{Ludwig}},\ }%
  \bibfield{journal}{%
  \bibinfo {journal} {Phys. Rev. B}\ }%
  \textbf{\bibinfo {volume} {78}},\ \bibinfo {pages} {195125} (\bibinfo {year}
  {2008})%
  \bibAnnoteFile{NoStop}{schnyder2008A}%
\bibitem{qi2008B}%
  \BibitemOpen
  \bibfield{author}{%
  \bibinfo {author} {\bibfnamefont{X.-L.}\ \bibnamefont{Qi}}, \bibinfo {author}
  {\bibfnamefont{T.}~\bibnamefont{Hughes}},\ and\ \bibinfo {author}
  {\bibfnamefont{S.-C.}\ \bibnamefont{Zhang}},\ }%
  \bibfield{journal}{%
  \bibinfo {journal} {Phys. Rev. B}\ }%
  \textbf{\bibinfo {volume} {78}},\ \bibinfo {pages} {195424} (\bibinfo {year}
  {2008})%
  \bibAnnoteFile{NoStop}{qi2008B}%
\bibitem{kitaev2009}%
  \BibitemOpen
  \bibfield{author}{%
  \bibinfo {author} {\bibfnamefont{A.}~\bibnamefont{Kitaev}},\ }%
  \bibfield{journal}{%
  \bibinfo {journal} {AIP Conf. Proc.}\ }%
  \textbf{\bibinfo {volume} {1134}},\ \bibinfo {pages} {22} (\bibinfo {year}
  {2009})%
  \bibAnnoteFile{NoStop}{kitaev2009}%
\bibitem{bender1998}%
  \BibitemOpen
  \bibfield{author}{%
  \bibinfo {author} {\bibfnamefont{C.~M.}\ \bibnamefont{Bender}}\ and\ \bibinfo
  {author} {\bibfnamefont{S.}~\bibnamefont{Boettcher}},\ }%
  \bibfield{journal}{%
  \bibinfo {journal} {Phys. Rev. Lett.}\ }%
  \textbf{\bibinfo {volume} {80}},\ \bibinfo {pages} {5243} (\bibinfo {year}
  {1998})%
  \bibAnnoteFile{NoStop}{bender1998}%
\bibitem{bender2007}%
  \BibitemOpen
  \bibfield{author}{%
  \bibinfo {author} {\bibfnamefont{C.~M.}\ \bibnamefont{Bender}},\ }%
  \bibfield{journal}{%
  \bibinfo {journal} {Rep. Prog. Phys.}\ }%
  \textbf{\bibinfo {volume} {70}},\ \bibinfo {pages} {947} (\bibinfo {year}
  {2007})%
  \bibAnnoteFile{NoStop}{bender2007}%
\bibitem{bender2010}%
  \BibitemOpen
  \bibfield{author}{%
  \bibinfo {author} {\bibfnamefont{C.~M.}\ \bibnamefont{Bender}}\ and\ \bibinfo
  {author} {\bibfnamefont{P.~D.}\ \bibnamefont{Mannheim}},\ }%
  \bibfield{journal}{%
  \bibinfo {journal} {Phys. Lett. A}\ }%
  \textbf{\bibinfo {volume} {374}},\ \bibinfo {pages} {1616} (\bibinfo {year}
  {2010})%
  \bibAnnoteFile{NoStop}{bender2010}%
\bibitem{jonessmith2010}%
  \BibitemOpen
  \bibfield{author}{%
  \bibinfo {author} {\bibfnamefont{K.}~\bibnamefont{Jones-Smith}}\ and\
  \bibinfo {author} {\bibfnamefont{H.}~\bibnamefont{Mathur}},\ }%
  \bibfield{journal}{%
  \bibinfo {journal} {Phys. Rev. A}\ }%
  \textbf{\bibinfo {volume} {82}},\ \bibinfo {pages} {042101} (\bibinfo {year}
  {2010})%
  \bibAnnoteFile{NoStop}{jonessmith2010}%
\bibitem{sinha2005}%
  \BibitemOpen
  \bibfield{author}{%
  \bibinfo {author} {\bibfnamefont{A.}~\bibnamefont{Sinha}}\ and\ \bibinfo
  {author} {\bibfnamefont{P.}~\bibnamefont{Roy}},\ }%
  \bibfield{journal}{%
  \bibinfo {journal} {Mod. Phys. Lett. A}\ }%
  \textbf{\bibinfo {volume} {20}},\ \bibinfo {pages} {2377} (\bibinfo {year}
  {2005})%
  \bibAnnoteFile{NoStop}{sinha2005}%
\bibitem{haldane1988}%
  \BibitemOpen
  \bibfield{author}{%
  \bibinfo {author} {\bibfnamefont{F.~D.~M.}\ \bibnamefont{Haldane}},\ }%
  \bibfield{journal}{%
  \bibinfo {journal} {Phys. Rev. Lett.}\ }%
  \textbf{\bibinfo {volume} {61}},\ \bibinfo {pages} {2015} (\bibinfo {year}
  {1988})%
  \bibAnnoteFile{NoStop}{haldane1988}%
\bibitem{Jackiw1976}%
  \BibitemOpen
  \bibfield{author}{%
  \bibinfo {author} {\bibfnamefont{R.}~\bibnamefont{Jackiw}}\ and\ \bibinfo
  {author} {\bibfnamefont{C.}~\bibnamefont{Rebbi}},\ }%
  \bibfield{journal}{%
  \bibinfo {journal} {Phys. Rev. D}\ }%
  \textbf{\bibinfo {volume} {13}},\ \bibinfo {pages} {3398} (\bibinfo {year}
  {1976})%
  \bibAnnoteFile{NoStop}{Jackiw1976}%
\bibitem{turner2010A}%
  \BibitemOpen
  \bibfield{author}{%
  \bibinfo {author} {\bibfnamefont{A.~M.}\ \bibnamefont{Turner}}, \bibinfo
  {author} {\bibfnamefont{Y.}~\bibnamefont{Zhang}},\ and\ \bibinfo {author}
  {\bibfnamefont{A.}~\bibnamefont{Vishwanath}},\ }%
  \bibfield{journal}{%
  \bibinfo {journal} {Phys. Rev. B}\ }%
  \textbf{\bibinfo {volume} {82}},\ \bibinfo {pages} {241102R} (\bibinfo {year}
  {2010})%
  \bibAnnoteFile{NoStop}{turner2010A}%
\bibitem{hughes2011}%
  \BibitemOpen
  \bibfield{author}{%
  \bibinfo {author} {\bibfnamefont{T.~L.}\ \bibnamefont{Hughes}}, \bibinfo
  {author} {\bibfnamefont{E.}~\bibnamefont{Prodan}},\ and\ \bibinfo {author}
  {\bibfnamefont{B.~A.}\ \bibnamefont{Bernevig}},\ }%
  \bibfield{journal}{%
  \bibinfo {journal} {Phys. Rev. B}\ }%
  \textbf{\bibinfo {volume} {83}},\ \bibinfo {pages} {245132} (\bibinfo {year}
  {2011})%
  \bibAnnoteFile{NoStop}{hughes2011}%
\bibitem{turner2010B}%
  \BibitemOpen
  \bibfield{author}{%
  \bibinfo {author} {\bibfnamefont{A.~M.}\ \bibnamefont{Turner}}, \bibinfo
  {author} {\bibfnamefont{Y.}~\bibnamefont{Zhang}}, \bibinfo {author}
  {\bibfnamefont{R.~S.~K.}\ \bibnamefont{Mong}},\ and\ \bibinfo {author}
  {\bibfnamefont{A.}~\bibnamefont{Vishwanath}},\ }%
  \bibinfo {howpublished} {arxiv:1010.4335}%
  \bibAnnoteFile{NoStop}{turner2010B}%
\bibitem{fu2011}%
  \BibitemOpen
  \bibfield{author}{%
  \bibinfo {author} {\bibfnamefont{L.}~\bibnamefont{Fu}},\ }%
  \bibfield{journal}{%
  \bibinfo {journal} {Phys. Rev. Lett.}\ }%
  \textbf{\bibinfo {volume} {106}},\ \bibinfo {pages} {106802} (\bibinfo {year}
  {2011})%
  \bibAnnoteFile{NoStop}{fu2011}%
\end{thebibliography}%
\end{document}